# Instructor Framing and Incentives Shape Physics Students' Engagement and Learning Gains from an Inquiry-Based Electrostatics Tutorial on the Method of Images

**Jaya Shivangani Kashyap \*, Robert P. Devaty and Chandralekha Singh**

Department of Physics and Astronomy, University of Pittsburgh, Pittsburgh, PA 15260, USA;
devaty@pitt.edu (R.P.D.); clsingh@pitt.edu (C.S.)
*Correspondence: jak509@pitt.edu

**Abstract**

The method of images (MoI) is a valuable technique for solving certain electrostatic boundary value problems consisting of charge density near conductor(s). We developed and validated an inquiry-based tutorial on MoI to help students learn to identify the problems involving symmetry in which MoI is applicable and then apply it by finding the correct image charge configuration. We implemented the inquiry-based tutorial accompanied by pretest and posttest, across three instructors' classes to evaluate student learning. We also conducted think-aloud interviews with advanced physics students, which helped us gain insights into their problem-solving strategies, evaluate their understanding developed through the tutorial and make necessary refinements to the MoI tutorial. The study identified common student difficulties, which were subsequently integrated into the inquiry-based tutorial as a guide to provide support to students. One important finding is that advanced students have common difficulties related to physics concepts similar to those found in introductory physics courses. The performance difference in the pretest administered after lecture-based instruction and the posttest administered after working through the tutorial reflects students' ability to apply what they learned from the inquiry-based tutorial compared to traditional lecture. Another important and unanticipated finding of this study is the potential impact of the framing of the inquiry-based tutorial and accompanying tests by one of the instructors on the engagement and performance of students. In particular, the instructor of one of the classes offered students a small amount of extra credit for engaging with the inquiry-based tutorial and tests, explicitly noting that these activities were not part of the current course syllabus and were primarily conducted to support physics education research. This kind of framing likely influenced students' motivation and engagement, which underscores how the way the instructor frames the inquiry-based instructional tasks to their students can have a significant impact on student performance. Overall, this iterative multi-year design-based comparative research with mixed-method triangulation provides valuable insights on the challenges involved in such studies that educators and researchers alike can greatly benefit from.

**Keywords:** tutorial; electrostatics; method of images; physics; scaffolding; problem-solving; instructor framing; research-based learning tools; incentive to learn; inquiry-based tutorial

# 1. Introduction and Framework

Developing a solid understanding of electricity and magnetism (E&M), including related aspects such as electrostatics, magnetostatics and electrodynamics, remains a persistent challenge conceptually and mathematically for students at both the introductory and advanced levels. For example, students continue to face obstacles in effectively applying underlying physics principles, a concern consistently reported in research on electricity and magnetism (Ding et al., 2006; Engelhardt & Beichner, 2004; Eylon & Ganiel, 1990; Gladding et al., 2012; Itza-Ortiz, Rebello, & Zollman, 2004; Lenaerts et al., 2002; Maloney et al., 2001; Savelsbergh et al., 2011; Suárez et al., 2024; Thacker et al., 1999; van Kampen & De Cock, 2023). Research has shown that students struggle in electrostatics with core concepts such as the electric field and the superposition principle (Campos et al., 2021; Campos et al., 2019; Li et al., 2023; Li & Singh, 2017, 2019). Furthermore, prior research in electrostatics has also elucidated the challenges faced by students in understanding conductors, insulators, capacitors, electric flux and Gauss' law in different contexts of learning about electrostatics (Ding et al., 2021; Galili & Bar, 2008; Guisasola et al., 2010; Li & Singh, 2018; Maries et al., 2017; Santana et al., 2023). Moreover, Faraday's law, electromotive force, and the general concept of electromagnetic induction (Garzón et al., 2014; Zuza et al., 2012, 2014, 2016) have also previously been discussed as being challenging for students. Prior work has also highlighted the importance of scaffolding, e.g., via guided problem solving, multimedia materials, computer simulation and kinesthetic teaching methods, in helping students develop a more coherent understanding of these concepts (Debowska et al., 2013; Goldberg & Otero, 2001; Richards, 2020).

Students in physics courses, including E&M, are expected to be proficient in mathematics, including trigonometry, vector calculus, differential equations, etc., and integrate them with physics concepts in diverse contexts (Bollen et al., 2016; Bollen et al., 2017; Bollen et al., 2015, 2018; Doughty et al., 2014; Nguyen & Rebello, 2011; Ryan et al., 2018; Wilcox & Corsiglia, 2019). In physics problem solving (Dufresne et al., 2005; Maries & Singh, 2023; T. J. Nokes-Malach & Mestre, 2013; Reif, 2008; Schwartz et al., 2005; Singh et al., 2023), mathematics serves not merely as a tool for computation but as a foundational element in students' interpretation and understanding of the physical system and in figuring out the solution approach that would be most productive for a given problem. Students often struggle with appropriately integrating physical and mathematical concepts while they are still building their expertise (Bing & Redish, 2007; Bing & Redish, 2009, 2012; Redish, 2004; Tuminaro & Redish, 2007). This can lead to difficulties in accessing the necessary mathematical knowledge they learned in prior courses or in integrating mathematics with conceptual reasoning in physics if appropriate scaffolding support is not provided (Dominguez et al., 2024; Galili, 2018; Justice et al., 2025; Karam, 2014, 2015; Karam et al., 2019; Keebaugh et al., 2024; Palmgren & Rasa, 2024; Pietrocola, 2008; Pospiech et al., 2019). Uhden et al. (Uhden et al., 2012) described technical and structural roles as two distinct roles of mathematics in physics. Specifically, "the technical skills are associated with pure mathematical manipulations whereas the structural skills are related to the capacity of employing mathematical knowledge for structuring physical situations" (Uhden et al., 2012). This framework is consistent with the perspective offered by Tzanakis (Tzanakis, 2016), who states that "mathematics is the language of physics, not only as a tool for expressing, handling and developing logically physical concepts, methods, and theories, but also as an indispensable, formative characteristic that shapes them, by deepening, sharpening, and extending their meaning, or even endowing them with meaning". In line with Uhden et al. (Uhden et al., 2012), we recognize the structural role of mathematics in physics. The structural role is particularly important, as mathematics is so highly intertwined with physics concepts that it is often difficult to separate them from each other. Furthermore, research indicates that physical and mathematical knowledge is closely interconnected, and students' epistemological framing plays a key role in integrating these domains while solving physics problems (DiSessa, 1993, 2018; Hammer, 1994; Odden & Russ, 2018; Redish, 2024; Russ & Odden, 2017; Schoenfeld, 2016; Sikorski & Hammer, 2017; Sirnoorkar et al., 2023).

Since a strong interdependence exists between math and physics, examining the difficulties students face while learning the concepts and solving problems in E&M and providing effective support, e.g., using inquiry-based approaches, to address the difficulties is very important. Researchers have undertaken substantial efforts to enhance student understanding of E&M concepts not only for introductory students but also for advanced physics students. For instance, Singh (Singh, 2006) used a research-based assessment tool to study how students from introductory to graduate level reason about symmetry and Gauss's law. Similarly, Bilak and Singh (Bilak & Singh, 2007) investigated students' understanding of conductors and insulators across both introductory and graduate level courses. A junior-level E&M course has been redesigned by Chasteen et al. (Chasteen et al., 2012b) and they also developed a diagnostic instrument to assess upper-division students' understanding of concepts in electrostatics (Chasteen et al., 2012a), which has also been employed at other institutions to evaluate learning outcomes for these topics (Zwolak & Manogue, 2015). Pepper et al. (Pepper et al., 2012) analyzed the specific challenges encountered by students while implementing mathematical tools in upper-level E&M contexts. Ryan et al. (Ryan et al., 2018) focused on student difficulties with boundary conditions in the context of electromagnetic waves, while Wilcox and Corsiglia (Wilcox & Corsiglia, 2019) explored how students struggle to integrate concepts from magnetostatics and classical mechanics. Bollen et al. (Bollen et al., 2015) investigated student challenges with vector calculus in electrodynamics, including how students apply divergence and curl (Bollen et al., 2016), and how the symbolic and graphical representations of vector fields (Bollen et al., 2017) are interpreted by them. In order to improve the understanding of vector calculus concepts in electrodynamics (Bollen et al., 2018), building on these findings, Bollen and colleagues also designed and evaluated a guided-inquiry teaching–learning sequence. Singh and Maries (Singh & Maries, 2013) studied how well graduate students in the core E&M course understand divergence and curl which is central for success in upper-level E&M. In another study, Mason et al. (Mason et al., 2023) examined performance gains in upper-division E&M when explicit incentives were provided to students to correct their mistakes. In addition to an inquiry-based tutorial on Laplace's equation (Kashyap et al., 2025), in two case studies, advanced students' sensemaking while solving physics problems in individual think-aloud interviews has been investigated in the context of Laplace's equation (Kashyap & Singh, 2025) and method of images (Kashyap & Singh, 2024, 2026) through the epistemic game framework.

Building on the previous research in physics education, this study focuses on the development, validation and in-class evaluation of an inquiry-based tutorial to provide support to students in learning the method of images. Inquiry-based tutorials in introductory physics have been developed and used extensively, e.g., by the group at the University of Washington, to improve student learning (McDermott, 1984, 2001, 2021; McDermott & Redish, 1999; McDermott & Shaffer, 1992; McDermott & Shaffer, 2002; Pride et al., 1998; Shaffer & McDermott, 1992). The method of images (MoI) is a valuable approach in electrostatics to obtain the potential for certain kinds of boundary value problems, e.g., involving charge distribution near conductors. The method is obtained from the corollary of the first uniqueness theorem, which states (Griffiths, 2017) that the potential in a volume $\mathcal{V}$ is uniquely determined if the charge density throughout the region and the value of the potential $V$ is known on all the boundaries. Taking advantage of this method, the original complex problem can be replaced by a different configuration containing image charges such that the charge density and boundary conditions in the desired region (region of interest) where the potential is to be calculated remain unchanged. This can be done by removing the grounded conductor and placing image charges in the excluded region (region where the original charges are not present). Students learn this technique in upper-level E&M courses. We find that students often have difficulty in recognizing when MoI can be effectively used. Furthermore, we find that even when students correctly identified that the method is applicable for a given configuration, many struggled, e.g., to determine the correct number, positions, magnitudes, and signs of the image charges.

Scaffolding plays a central role in helping students navigate complex physics problems (Chaiklin, 2003; Lin & Singh, 2015; McLeod, 2010). Scaffolding refers to the instructional support provided to students as they engage with challenging tasks, e.g., in an inquiry-based teaching-learning sequence in the tutorial, with the goal of gradually building their content knowledge and skills for solving such problems independently. Vygotsky's Zone of Proximal Development (ZPD) (Vygotsky, 1978) describes the space between what learners can accomplish on their own and what they can do with guidance from a more knowledgeable person or in collaboration with others. Learning is most productive when instruction targets this space, which changes as learners grow in their understanding. For inquiry-based instruction to be effective, it should connect with what students already know and build upon it. In physics education, this support for inquiry-based learning often consists of decomposing complex problems into smaller sub-problems, providing helpful prompts or inquiry-based sequences throughout. When implemented thoughtfully, these inquiry-based strategies help students progress from basic understanding to more advanced levels of problem solving. Scaffolding can help in reducing the cognitive demands of the task, help students focus their attention on key ideas, and promote reflection and self-regulation (Ding et al., 2011). Well-designed learning activities, such as the guided inquiry-based tasks involving the method of images, can help students operate within and expand their ZPD (Vygotsky, 1978). In particular, keeping in mind students' ZPD, scaffolding can support their learning and improve their ability to solve complex physics problems.

Based on the theoretical framework of Vygotsky's Zone of Proximal Development (ZPD) (Vygotsky, 1978), we developed and validated an inquiry-based tutorial and corresponding tests on the method of images aimed at scaffolding student learning and helping them build expertise in solving problems. In addition to individual student interviews and expert feedback during development and validation of the tutorial, the tutorial and corresponding tests were implemented in class over four consecutive years in three instructors' classes. The theoretical, methodological and practical contributions of this study highlight the challenges in a design-based comparative study involving quasi-experimental design and mixed-method triangulation during the development, validation and implementation of research-based learning tools. The study provides insight into common student difficulties in understanding the method of images and shows how these difficulties persist across different instructional contexts and student levels. The study also reveals the potential value of instructor framing on student engagement and learning. For example, the way the third instructor, who administered the latest version of the tutorial with the most effective scaffolding based upon our out-of-class interview findings, framed the activity to students could have impacted student engagement and learning with research-based tools. In particular, that instructor framed these activities as something not relevant for the course and intended primarily to support education researchers, which may have reduced student motivation to engage with the tutorial and corresponding assessments. When students view an inquiry-based activity as valuable and connected to the course goals, they are more likely to participate meaningfully. These findings can be useful resources for those who are interested in conducting similar design-based studies.

## 2. Methodology

### 2.1. Development and Validation

To scaffold student learning and support them in developing expertise in solving electrostatics problems using the MoI when the method is effective, we investigated student difficulties and developed a research-based, guided inquiry-based tutorial along with the corresponding tests. The tutorial is designed to provide scaffolding support and help students deepen their understanding of what the method entails. The learning sequences in the inquiry-based tutorial provide support to help students recognize the conditions under which the MoI can be useful as well as the importance of checking the boundary conditions to ensure the correctness of the image charges.

We note that this study is a design-based research study (A. L. Brown, 1992; Collins et al., 2004) and the development and validation of the tutorial and the corresponding pre/posttests is an iterative design process. Due to the complexity and diversity of students and instructors in real classrooms, classroom interventions are different from lab studies outside of class (e.g., individual student interviews conducted in this study outside of class to understand student thought processes as they solve problems and work through a particular version of the tutorial). The theoretical and methodological challenges in conducting design experiments to improve learning in classroom settings were laid out elaborately by Brown and Collins (A. L. Brown, 1992; Collins et al., 2004) and their accounts about design-based research in the classroom make it clear that researchers must iterate classroom interventions based on the outcomes of previous implementations to improve student learning. Thus, despite the initial development and content validation of the tutorial and pre/posttests outside of class via expert feedback and individual student interviews, the student outcomes based on real classroom implementation over a four-year period in three different instructors' classes led to additional iteration of the learning tools. We note that we use a mixed-method triangulation involving both individual interviews and in-class administration throughout the development and validation process.

In this quasi-experimental design study, the development and validation of the inquiry-based tutorial and the corresponding pre-/posttests started with a cognitive task analysis from the perspective of an expert. It included an investigation of common difficulties students encounter while learning about the MoI for solving electrostatics problems in an upper-level course in E&M. Using common difficulties students experienced in solving the problems that can be effectively solved using the MoI as a guide, we developed a guided inquiry-based tutorial via an iterative process. The tutorial leverages students' relevant prior knowledge and difficulties that emerged during student problem-solving over several years, based on homework submissions, exams, and interviews. Discussions among researchers and other subject-matter experts, particularly upper-level E&M instructors throughout, provided valuable insights and played an important role in the development and content validity of the inquiry-based MoI tutorial and corresponding tests. The guided inquiry-based sequences in the tutorial provide students with opportunities to learn when the MoI is valuable and use it to solve problems by offering adequate scaffolding support and feedback.

After formulating the tutorial's learning objectives and the corresponding tests and leveraging insights from students' difficulties found via research, the tutorial, pretest and posttest were developed. The pretest is used to evaluate student learning after traditional lecture-based instruction on the MoI. The inquiry-based tutorial strives to help students develop the ability to approach problems in an expert-like manner by providing scaffolding and guidance. The posttest after the tutorial evaluates the learning of students after engaging with the inquiry-based sequences in the tutorial. A comparison of the performance of students on the pretest and posttest helps us understand the effectiveness of the tutorial. We conducted semi-structured, think-aloud, online individual interviews to understand the rationale behind students' learning before, during, and after the development of different versions of the inquiry-based sequences in the tutorial and the corresponding pretest and posttest.

The researchers used a thematic analysis to organize and interpret interview data (Braun & Clarke, 2006, 2019) with the tutorial and pre-/posttest questions themselves serving as pre-defined overarching themes. Our interview coding is structural which is a holistic approach based on the problem features (McLellan-Lemal & MacQueen, 2008; Saldana, 2021). Two of the authors collaboratively reviewed the interviews. After multiple iterations, both researchers discussed and agreed on the analysis of student difficulties. Individual interviews helped us refine the tutorial as well as the tests and make changes to improve student learning and also helped in content validity, to make sure the content is interpreted as intended. These efforts also contributed to reducing student difficulties and strengthening the effectiveness of the inquiry-based sequences in the tutorial. Iterative improvements and revisions were made throughout the process using collaborative discussions among the three authors, supplemented by insights and feedback from other course instructors and subject-matter experts for content validity.

Both unscaffolded and scaffolded tests (Test A and Test B) were developed and validated along with the tutorial. The study involved two versions of scaffolded tests, Test A and Test B (latest version of Test A is included in Appendix A), in different years of administration in the classroom using a convenience sampling method. In each year, one of the tests was used as the pretest and the other as the posttest. The unscaffolded tests were in the format of a typical textbook question with only relevant information given in the problem without guidance (i.e., the problem was not broken down into subproblems). On the other hand, the scaffolded tests (Test A and Test B) break down the problem into sub-problems and include prompts designed to provide scaffolding support. The scaffolded pre-/posttests each consist of six questions based on a problem with a charge and grounded conductor(s). Test B included a problem that is somewhat more challenging with six-fold symmetry than Test A (with two image charges present near a grounded conducting plane). We initially kept Test B more challenging for the posttest after students engaged with the tutorial, consistent with the approach used in tutorials developed by the University of Washington to test if students are able to perform more challenging posttests (compared to pretests) after engaging with the tutorial (McDermott & Shaffer, 2018). Students' ability to explain the purpose of the method of images and its relation to the uniqueness theorem, as well as the necessary and sufficient conditions for the solution to be unique (uniqueness theorem to hold), was evaluated using scaffolded tests. The tests evaluated students' ability to identify the number of image charges for a given configuration, boundary conditions, whether the conductor should be removed when image charges are placed, and to find the electrostatic potential in the region of interest.

In the scaffolded version of the tests, three of the questions in Test A and Test B, Q1, Q2, and Q5, were identical in the pretest and posttest, and especially focused on the conceptual understanding of the foundational issues related to MoI. Q3, Q4, and Q6 can be answered based on the given configuration of charge and grounded conducting plane(s). Based on the feedback from some of the instructors, we later added a figure in the problem posed in Test B; however, this small change did not produce any significant difference in student performance. In the last year of implementation, we revised the inquiry-based tutorial by incorporating an additional twelve-fold symmetry problem (in which space can be conveniently divided into twelve equal parts based upon the number and locations of point changes), in which students were asked to identify the correct number of image charges, along with the hint provided to guide them through the correct answer. An additional short problem, Q7 in Test A on pretest, was added (only asking students to identify the number of image charges) based on the twelve-fold symmetry problem (the same problem that was scaffolded in the tutorial). The purpose of adding this problem in Test A on pretest was to evaluate the extent to which students could take advantage of the scaffolding provided in the tutorial and improve their performance on Test B on posttest (on a six-fold symmetry problem) after engaging with the tutorial. While students' ability to identify whether the method of images is suitable and apply it for calculating the potential is measured by both the unscaffolded and scaffolded pretest and posttest in interviews, only the scaffolded pretest and posttest were used in in-class administration of the tutorial and accompanying pre/posttests.

The Cronbach's alpha for posttest ranged between 0.6 and 0.7 for three instructors, indicating moderate internal consistency (Raykov & Marcoulides, 2019). Since the tests evaluate different aspects of understanding of the method of images, such as the ability to explain what the method is, identify the location and number of image charges, write the boundary conditions for a given configuration mathematically, write the expression for the electric potential, etc., one should not necessarily expect Cronbach's alpha to be high. Students could differ in understanding different aspects of the MoI concepts. Other researchers in physics education have also noted that Cronbach's alpha values can be low for a test (McKagan et al., 2010). These authors (McKagan et al., 2010) state that "It is commonly stated that [Cronbach's alpha] should be above the cut-off value of 0.7 for an instrument to be reliable. However, Adams and Wieman (Adams & Wieman, 2011) argue that instruments designed to measure multiple concepts may have low Cronbach Alpha because these concepts may be independent". Additionally, since all three instructors have similar values for Cronbach's alpha, the test appears to function similarly across instructional contexts.

To investigate differences in pretest and posttest, we calculated effect sizes and normalized gains since both of these measures are considered useful (Nissen et al., 2018).

### 2.2. Inquiry-Based Tutorial Implementation in the Course and Details of Instructors

The tutorial was administered in-class in an upper-level electromagnetism (E&M) course at a large research-intensive university in the US. At this university, an upper-level E&M course is offered as a two-semester course sequence. The first-semester course covers multiple approaches to calculating electrostatic potential, including the method of images (MoI). The prerequisites to enroll in the course are introductory physics 2 (E&M) and Calculus 3, which involves vector calculus. Additionally, a mathematics course in differential equations is a co-requisite.

Once satisfactory initial versions of the tests (Test A and Test B) and the inquiry-based tutorial (Tutorial) were developed, they were implemented over two years in instructor 1's (I1's) classroom, following traditional lecture-based instruction on the topic. The tutorial was subsequently introduced in the courses of two additional instructors, instructor 2 (I2) and instructor 3 (I3).

The tutorial, along with the pretest and posttest, was administered, following traditional lecture-based instruction on the topic, over four consecutive academic years by three instructors: I1, I2, and I3. Although students would likely benefit most from engaging with the inquiry-based sequences in the MoI tutorial in small, instructor-guided groups during class, all three instructors chose to assign it as homework to preserve class time. Throughout all four years of in-class administration, instruction was based on the textbook *Introduction to Electrodynamics* by Griffiths. Over four years, the upper-level E&M course was taught by three instructors: I1 for the first two years, I2 in the third year, and I3 in the fourth year, who taught the spring-semester second course of the sequence. The inquiry-based tutorial was assigned at the beginning of I3's course, following students' traditional lecture-based instruction on the material in the first-semester course.

Following I1's preference, students received the pretest and posttest in a scaffolded format. To ensure consistency across all instructors, the same scaffolded format was used for the pretest and posttest administered by I2 and I3 as well. Figure 1 provides a depiction of the chronological order in which different components were administered in class.

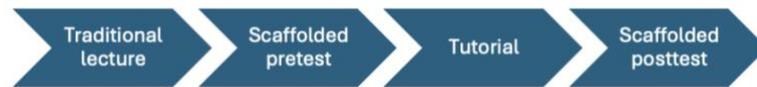

**Figure 1.** Chronological order of in-class implementation of the set of pretest, inquiry-based tutorial and posttest.

For I1 and I3, Test A was given as a pretest and Test B was given as a posttest along with the tutorial. For I2, Test B was given as a pretest and Test A was given as a posttest along with the tutorial. The most recent version of the tests and the tutorial were given to students in I3's class and are also provided in the Supplementary Material.

I1 conducted the pretest and posttest as graded exams or in-class quizzes, while I2 offered them for extra credit. In I1's and I2's courses, the tutorial was assigned as homework after students had completed lecture-based instruction on the relevant topics. For I3, who taught the second course of the upper-level E&M course sequence, the pretest was administered in the first week, followed by the tutorial as homework and then the posttest. With respect to instructor framing, I3 explained to students that the tutorial and tests were part of an educational research study and were not tied to the course content, which covered electrodynamics rather than electrostatics in the second semester of the upper-level E&M sequence. Students were informed that they would receive a small amount of extra credit for completing the pretest, tutorial, and posttest, irrespective of correctness.

Table 1 summarizes the number of students who took each test (all students), along with the number of students who completed both the pretest and posttest (matched students) in in-class administration.

**Table 1.** Number of students completing the pretest, posttest, and both tests (matched students) in each instructor's class.

| | Number of Students | | |
|---|---|---|---|
| Instructor | Pretest (All) | Posttest (All) | Pretest and Posttest (Matched) |
| I1 | 50 | 52 | 50 |
| I2 | 17 | 19 | 12 |
| I3 | 15 | 17 | 15 |

Approximately 24% of the pretests and posttests were independently graded by two graders using the rubric in Appendix A to evaluate interrater agreement. The overall disagreement across questions ranged from 0% to 11%. This shows a good inter-rater agreement between different graders.

### 2.3. Interview Details

The development and validation of the preliminary version of the tutorial and tests involved identifying student difficulties through homework, exams, discussion with experts and multiple student interviews. Once the preliminary version of the tutorial and tests was ready, the tutorial, along with tests, was implemented in the classroom over a four-year period while further interviews with thirteen advanced students were conducted during that period. To enhance the inquiry-based sequences in the tutorial and address student difficulties, we made iterative revisions based on think-aloud interviews with students. Early discussions (Kashyap & Singh, 2026) during the inquiry-based sequences' development and validation helped identify common difficulties and informed strategies to support student understanding, making sure content is interpreted appropriately. Thirteen advanced students, including eight graduate and five undergraduate students, were interviewed at various stages. At this stage, the first round of think-aloud interviews involved six graduate students, who had previously taken the upper-level E&M course. A later round included five upper-level undergraduate students, who were either currently enrolled (and had received instruction on the relevant concepts) or had completed the course previously, along with two additional graduate students. While the inquiry-based tutorial targeted advanced undergraduate students, interviews with graduate students offered additional understanding of student learning and the effectiveness of inquiry-based learning sequences in the tutorial. No significant performance differences were found between graduate and undergraduate participants.

The interviews took place online using Zoom, and students received the pretest, inquiry-based tutorial, and posttest as PDF files. Students used electronic devices such as iPads/tablets to share their screens while scribbling on the PDF. They replaced their names with pseudonyms, kept the video off to keep their identities anonymous and shared the screen while solving the problems. In this paper, we use pseudonyms for the students and 'they/their' as pronouns to keep their gender identity undisclosed. Students shared their scribbled PDF file with the interviewer at the end of each session. These interviews were conducted using a semi-structured think-aloud protocol in which students were asked to speak out loud as they solved or thought about the solution of a given problem. The interviewer did not interrupt or help during the interview except when clarification was needed. These interviews were typically conducted in three different sessions on different days of the week with each student. Interviews were conducted with each student for an average of 3 h 20 min total across the three sessions. During interviews, to observe how effective the scaffolding was, we provided students with the unscaffolded test, and immediately after that, we provided them with the scaffolded test. Figure 2, shown below, gives a depiction of the chronological order in which different components were administered in interviews.

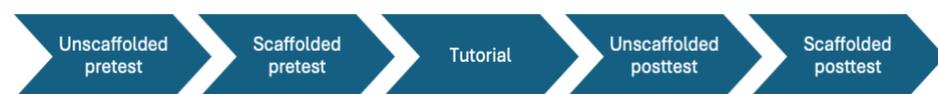

**Figure 2.** Chronological order in which tests and inquiry-based tutorial were administered to the student interviewee.

The tutorial, pretest and posttest were tweaked based on feedback from students during the interviews. In particular, we iterated the inquiry-based sequences in the tutorial and the tests several times based upon the feedback from interviewed students and physics experts including instructors of the upper-level E&M course. Thus, the inquiry-based tutorial on the method of images is the outcome of multiple iterations based on feedback at various stages, which helped fine-tune it and improve its flow. Figure 3 shows the design-based study involving mixed-method triangulation during the iterative process of development and validation of the tutorial and tests.

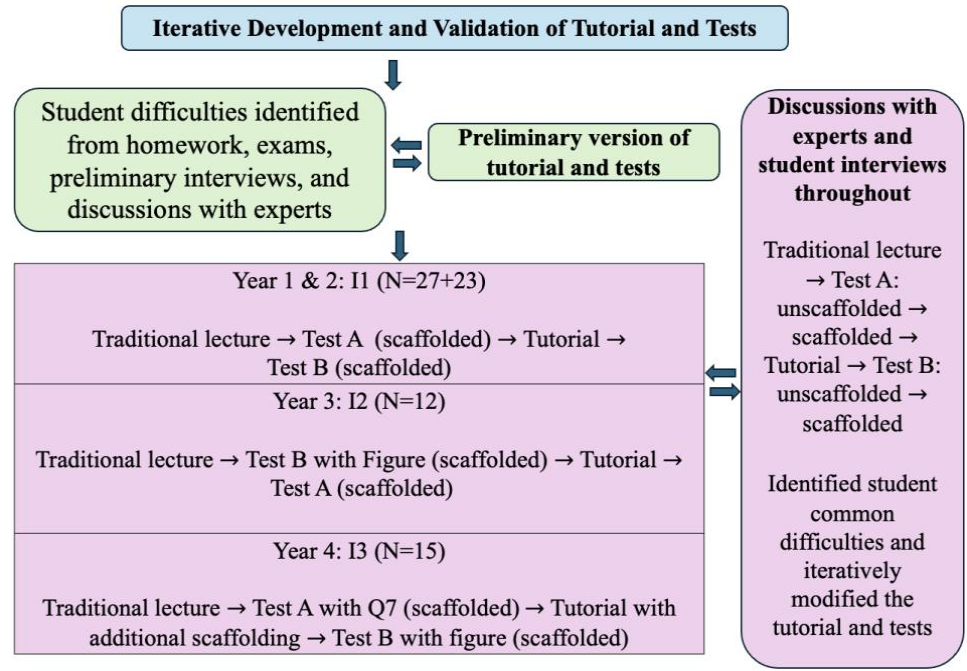

**Figure 3.** Illustration of the design-based study involving mixed-method triangulation during the iterative process of development and validation of the tutorial and tests.

### 2.4. Learning Objectives of the Inquiry-Based Tutorial

The tutorial on MoI provides scaffolding support to students and asks them to make a diagrammatic representation for the given configurations. It strives to help students learn when the MoI is an effective approach, how it is related to the uniqueness theorem and what aspects of the uniqueness theorem allow us to replace the conductor with the image charges. It strives to help students contemplate how the electric field lines in a given problem in the region of interest are identical in the two configurations (original problem with the conductor and the problem with image charges without the conductor). This comparison of electric field lines can be used to find the potential in the region of interest readily using the image charges. The inquiry-based sequences in the tutorial scaffold student learning and help them identify and check the boundary conditions to make sure the image charges are placed correctly. Then the tutorial helps students find the electrostatic potential in the region of interest with the help of the superposition principle using the expression for the potential due to point changes. Scaffolding support is provided for writing the distances in the expression since difficulties related to writing distances while finding the potential were identified in individual interviews.

Table 2 lists the learning objectives of the tutorial (Tutorial), alongside the corresponding pretest or posttest (Test A/Test B) and the inquiry-based tutorial questions.

**Table 2.** Learning objectives of the latest version of the inquiry-based MoI tutorial and the corresponding concepts evaluated in Test A and Test B. Each learning objective starts with "Students should be able to…".

| Learning Objectives | Test A/Test B (Latest Version) | Corresponding Tutorial Questions |
|---|---|---|
| draw the diagram for the given configuration | | 1a, 1c, 1d, 1f, 7a, 17a, 16c, 22 |
| explain how the induced charges are distributed on the conductor and how they contribute to the potential. | | 1b, 1e, 2 |
| explain the relevance of the uniqueness theorem to the method of images | 1, 2 | 3,4,13, 16a, 19 |
| identify the region of interest and excluded region in a given configuration | | 5a, 5b, 16b |
| draw the electric field lines in different regions and/or confirm that the potential in the image problem in the region of interest is the same as the method of images (after the image charges are placed and the conductor is removed) is the same as the potential due to the point charge and induced charges on the grounded conductor(s) | 5 | 5c, 5d, 8 |
| write/identify the boundary conditions | 4 | 6a, 6b, 7b, 7c |
| write expression for distances from each point charge to the point where the potential is to be calculated | | 9, 10,11 |
| use the superposition principle to find the potential due to various point charges | 6 | 12, 17d |
| calculate the induced charge density by taking the normal derivative of the potential | | 14 |
| identify that the total induced charge on the grounded plane is the same as the total image charge | | 15 |
| identify the number of image charges required for the given configuration | 3, 7 [1] | 17b, 20, 21 |
| check the boundary conditions | | 17e, 18 |

[1] This question was only added to Test A.

## 3. Research Questions

Here we focus on the following research questions.

RQ1: What were student difficulties observed during the development and validation of the tutorial along with the pretest and posttest?

RQ2: How did the tutorial help in reducing the student difficulties that were observed?

RQ3: Did additional scaffolding in the tutorial for a complex problem involving twelve-fold symmetry help in improving student performance on Test B on a six-fold symmetry problem in the posttest?

## 4. Results and Discussion

In this section, we describe the process of development and validation of the tutorial and tests as a quasi-experimental study involving design-based research. The results pertaining to all three research questions are also described in detail. We discuss the challenges students face in solving problems related to MoI and how the inquiry-based sequences in the tutorial were designed to support students in addressing those challenges.

### 4.1. Student Difficulties and the Role of Inquiry-Based Tutorial in Supporting Learning

We first describe the difficulties students had with the method of images during interviews. We then discuss how the inquiry-based MoI tutorial helped in addressing those challenges and how it was iteratively refined to better support student learning.

#### 4.1.1. Student Difficulty with the Excluded Region

Regarding RQ1, in the initial interviews, we observed that even when students correctly recognized in the scaffolded pretest and the inquiry-based tutorial that the method of images was the appropriate method and that they needed to determine the image charges, they often lacked a clear understanding of how to distinguish the excluded region from the region of interest for a given charge configuration. In particular, they were unsure about identifying the excluded region, where image charges must be placed, and how it is different from the region of interest, where the electric potential is calculated (and no image charge can be present). One student, for example, misinterpreted the configuration given in the problem in the scaffolded pretest and placed the given point charges in both the region of interest and the excluded region. Eventually, they ended up placing image charges in the regions where there were charges present. This error reflects a fundamental difficulty in visualizing and separating the excluded region from the region of interest for the given configuration.

Regarding RQ2, to help students develop a clearer understanding of how to define these regions and help them visualize how the solution in the region of interest is the same for the original problem and the problem with the correct image charges, two figures were added in a problem in the tutorial asking students to label the region of interest and the excluded region and draw electric field lines when a charge $+q$ was placed near an infinite grounded conducting plane in the first figure and when there was an electric dipole (with charges $+q$ and $-q$) in the second figure. They were asked the following questions:

(a) *Label the region of interest and the excluded region in the first figure (provided to students), which is based on our original problem.*

Students were then asked to reflect on a conversation between two hypothetical students, John and Emily, and consider whether they agree or disagree with each:

(b) *Consider the following conversation between John and Emily:*

*__John:__ The region of interest is the region where the potential is to be found. The value of the potential on all boundaries including at infinity of the region of interest is specified.*

*__Emily:__ I agree. Also, in the method of images, no image charge can be placed in the region of interest, i.e., image charges are only placed in excluded region, where we are not asked to calculate the potential.*

John and Emily are correct. In the hypothetical conversation, the students are given two additional problems in which they are asked to sketch the electric field lines in different regions. The problems are designed to help students recognize that the electric potential in the region of interest is the same whether the grounded conductor is present or replaced by image charges, demonstrating how the MoI works.

In the later part of the inquiry-based sequence in the tutorial, another problem encouraged reflection on the concepts they had learned so far. In this task, students were asked to explain whether the image charges in the modified configuration were in the region where the potential is to be calculated as follows:

*In our original problem, we wanted to find the potential in the space above the grounded conducting sheet (throughout the region of interest). In our new problem, we introduced a point charge, which we called an image charge, and removed the grounded conducting sheet from the problem. Was the image charge in the region in which we wanted to find the potential? Explain your reasoning. (Hint: Go back to the conditions for the uniqueness theorem and see if any of the conditions are violated if the image charge is placed in the region of interest.)*

A hint is provided encouraging students to revisit the conditions for the uniqueness theorem and consider whether placing the image charges in the region of interest would violate any of those conditions. Students demonstrated improved understanding of the concept after working through the tutorial.

For example, one of the interviewed students answered the given question by saying, '*...was the image charge in the region in which we wanted to find the potential … I guess I wasn't that careful about it. But no...new point charges are located in the excluded region. Okay that's important because that is basically what…I've been seeing this whole time in this part...so the bottom* [region] *was the excluded region, I think...to calculate the potential…the mirror or the image charge was always in the excluded region'*. Another student said, '*the new point charge* [image charge] *that we were adding was in the excluded region. So, we took this to be our region* *of interest* [region $z > 0$]. *So, our* [image] *point charge that we added was in…the excluded region …'.*

The instances suggest that the tutorial facilitated students' ability to differentiate between the region of interest and the excluded region.

### 4.1.2. Difficulties Experienced by Students in Interpreting Grounding

Regarding RQ1, in interviews, we observed that students often exhibited confusion related to the concept of grounding. In many cases, they interpreted 'grounded' as meaning 'neutral' or implying the absence of charge. For example, one interviewed student, while engaging with the tutorial, said, *'And it's grounded. So…conductors have a net charge of 0, especially if it's grounded.'.* Similarly, another student said, *'If it's a grounded conductor…then I think that surface charge is 0. Is that what I'm remembering?'* This student later changed their mind about the grounded conductor having zero surface charge.

Regarding RQ2, to help students with the concept of charge distribution on a conductor with and without grounding, in the inquiry-based tutorial, students are first asked to sketch the given configurations as follows:

(a)  *Let's consider a situation with which you may be familiar from an introductory E&M course: an isolated spherical conductor with a point charge +q outside it. Draw a diagram that illustrates qualitatively the induced charge distribution on the conductor due to +q.*

Building upon this problem, they were asked to show what happens if the conductor is grounded.

(b)  *Draw a diagram of the situation given in the previous problem if the isolated spherical conductor is grounded with a point charge +q present outside it.*

We also included a discussion between two hypothetical students, Emily and Meera, to help students reflect upon their solution, as follows. Students were asked to indicate their agreement or disagreement with the statements and explain their reasoning.

**Emily:** *In the sketch of (a), negative induced charges are attracted to +q and positive induced charges are repelled from +q so the induced charge distribution on the conductor is as shown on the left in Figure 4. In (b), there will be no charge left on the surface of the conductor as shown on the right in Figure 4 because it is grounded.*

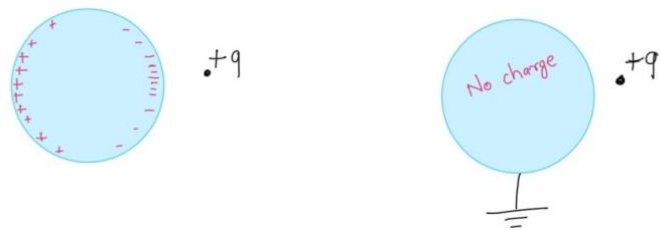

**Figure 4.** Diagram drawn by Emily showing the charge distribution (positive and negative charges) on an isolated spherical conductor with a point charge  +q  present outside it when (a) the conductor is not grounded (**left**) and (b) when it is grounded (**right**).

**Meera:** *I agree with your left picture in Figure 4 for (a), but I disagree with the right picture in Figure 4 for (b). Grounding a conductor does not necessarily mean it is neutral. When the conducting shell is grounded in the presence of a charge  +q near it, free electrons would flow from the earth to the conductor to neutralize the induced positive charges such that the potential at the surface is the same as that of the Earth. Upon grounding, there will be net negative charge on the conductor which will mostly be concentrated on the portion of the surface close to  +q.*

In this problem, Meera is correct. Meera also highlights a common difficulty students face, i.e., assuming that grounding means the conductor has no charge, and she helps clarify this situation. Checkpoints were later provided to help students reflect on how charges move between a conductor and the Earth when the conductor is grounded. We observed that these additions helped students build a clearer understanding of the concept of grounding.

### 4.1.3. Challenges in Identifying the Number, Sign, Magnitude, and Position of Image Charges

Regarding RQ1, another very common challenge in applying the method of images was recognizing the correct image charge configuration needed to determine the potential correctly. We found that many students had difficulty recognizing the symmetry of the configuration. For example, in a tutorial problem that required three image charges, several students initially claimed that only two image charges were sufficient. However, the structure of the inquiry-based sequences, e.g., which asked students to verify the boundary conditions later, helped them realize that three image charges were necessary. The configuration used in Test B, which involved six-fold symmetry (in which space can be conveniently divided into six equal parts based upon the number and locations of point changes), also proved to be challenging for many students. They had difficulty using the symmetry of the charge distribution effectively.

Regarding RQ2, to address this issue, we included a twelve-fold symmetry configuration in the tutorial in which students are asked to identify the correct number of image charges. For example, in the tutorial, a hypothetical student, Emily, shared a problem with her friends as follows:

*Consider two grounded conducting planes as shown in Figure 5, both perpendicular to the x-y plane, intersecting at an angle of 30°. One plane is at y = 0 and the other makes a 30° angle with respect to it. A point charge +q is placed at a point along the line bisecting the region at 15° in the region of interest.*

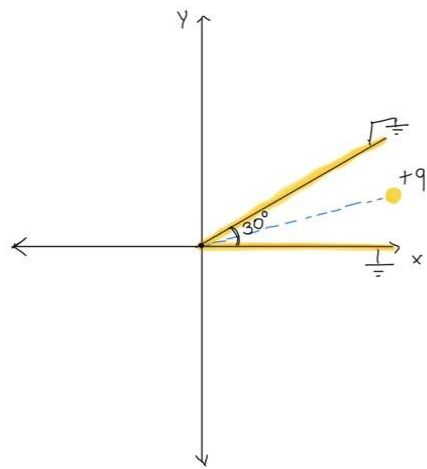

**Figure 5.** A point charge '*q*' present in the region of interest between two grounded conducting planes intersecting at 30°.

*How many image charges are required to solve the problem shared by Emily? Explain your reasoning.*

After this, students are asked to compare a previous problem from the tutorial involving two semi-infinite, perpendicular, grounded conducting planes with a point charge +q placed at x = +a (a > 0), y = +b (b > 0), z = 0 for the twelve-fold symmetry problem. Steve demonstrates a common difficulty in reasoning that was frequently observed during interviews. In contrast, another student, John, provides the correct reasoning by explaining how the symmetry of the image problem determines the required image charges as follows:

*Consider the following discussion between Steve and John regarding the problem shared by Emily:*

**Steve:** *In an earlier problem with two grounded conductors (semi-infinite, perpendicular, grounded conducting planes with a point charge $+q$ placed at $x = +a$ $(a > 0)$, $y = +b$ $(b > 0)$, $z = 0$), we found three image charges. Since this problem (twelve-fold symmetry) also has two grounded conducting planes, we get three image charges. But the positions of those image charges can be different based on the coordinates of the charge given in this problem.*

**John:** *I disagree, I think that in the earlier problem, the x–y plane was divided into four equal parts and hence the space in the problem had a four-fold symmetry. For the given situation, it is convenient to divide the space into twelve equal parts (obtaining twelve-fold symmetry) as shown in Figure 6.*

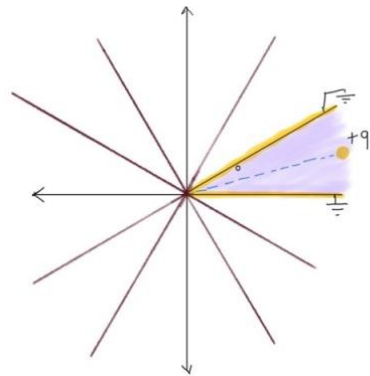

**Figure 6.** Picture drawn by John for the problem shared by Emily showing symmetry lines for the problem in which a point charge $+q$ is placed in the region of interest (shaded) between two grounded conducting planes placed at 30°.

Students were then given a hint suggesting that there are eleven image charges in total, but they were asked to decide whether they agree or disagree with this claim and justify their reasoning as follows:

*After placing 11 image charges symmetrically in Figure 6, we can write the expression for the potential and check that the boundary conditions are satisfied so that the potential in the new problem is zero on the surfaces where the grounded conducting planes were in the original problem.*

*Is John correct? Explain your reasoning*

John is correct, and the problem requires eleven image charges. Students are then asked to represent all eleven image charges in Figure 6.

*Assume that John wrote the expression for the potential correctly with 11 image charges and found that all the boundary conditions are satisfied. Place the image charges in Figure 6 drawn by John.*

This additional problem with 11 image charges gave students an opportunity to become familiar with this type of configuration, struggle productively, and spend time thinking through symmetry. Finally, when the six-fold symmetry problem was given in Test B as a posttest to the students, interviewed students appeared to benefit from the scaffolding in the tutorial with the problem involving 12-fold symmetry and showed improved performance in the posttest.

### 4.1.4. Difficulty in Defining the Electrostatic Potential at Any Arbitrary Point $(x, y, z)$ in the Region of Interest

Regarding RQ1, we observed that many students who managed to locate the image charges correctly/partially correctly still struggled with writing the electrostatic potential at any arbitrary point $(x, y, z)$ in the region of interest. A common issue involved the incorrect interpretation of the symbol '$r$' in the expression for potential due to a point charge $\left(V = \frac{kq}{r}\right)$. Some students appeared to treat '$r$' in the expression of the potential as the distance of a point charge on the $z$-axis (in a given problem) to the origin, rather than as the distance between the point charge and an arbitrary point $(x, y, z)$ in the region of interest where the potential is to be evaluated. Those students also did not explicitly mention that they were determining the potential at the origin, and they chose '$r$' as the distance of the point charge on the $z$-axis from the origin. This issue reflects a physics-related difficulty in that the students struggle to recognize that when the potential is to be found in the region of interest, it must be found at an arbitrary point in that region and expressed as a function of $(x, y, z)$. In particular, the contribution from each point charge depends on the distance from that charge to the point in space where the potential is being calculated. A second issue is related to mathematics as students had difficulty in correctly expressing the distance between two arbitrary points. For example, one of the students in the scaffolded version of the pretest (Q6, Test A, given in the Appendix A and Supplementary Material) correctly recognized that the method of images should be applied to determine the electric potential for the given configuration but placed one of the given charges incorrectly in the excluded region. This may have happened due to misinterpretation of the problem statement. They correctly placed one of the given charges $+q$ at a distance '$a$' from the origin on the positive $z$-axis but incorrectly placed the other charge $-2q$ at the distance '$b$' from the origin on the negative $z$-axis (instead of placing it on the positive $z$-axis). To apply the method of images, they further placed a $+q$ image charge at a distance '$a$' from the origin on the negative $z$-axis and a $-2q$ image charge at a distance '$b$' from the origin on the positive $z$-axis. Not only did they apply the method of images incorrectly, but they also did not show correct interpretation of '$r$' in the expression of potential. They added the electric potential from each point charge as (ignoring the multiplicative constant), $\frac{q}{a}$ (given charge '$q$' at a distance '$a$' on the positive $z$-axis), $-\frac{2q}{b}$ (image charge '$-2q$' at a distance '$b$' on the positive $z$-axis), $\frac{q}{a}$ (image charge '$q$' at a distance '$a$' on the negative $z$-axis), and $-\frac{2q}{b}$ (given charge '$-2q$' at a distance '$b$' on the negative $z$-axis). Based on the charge configuration and distances they had, they wrote the expression for the potential due to point charges as $V = \frac{q}{a} - \frac{2q}{b} + \frac{q}{a} - \frac{2q}{b}$. This expression shows that along with the incorrect implementation of the method of images, they did not write distances correctly in the expression of the potential at any arbitrary point in the region of interest.

Similarly, another interviewed student in the scaffolded pretest (Q6, Test A) identified that two image charges were needed in addition to the two given charges and placed those image charges correctly in the excluded region. While discussing boundary conditions, they mentioned that the potential is zero at the surface of the conductor and at infinity (they assumed the conductor to be a thin slab with its thickness tending to zero), so the only regions where the potential can be non-zero are the regions outside the conductor. They correctly placed two given charges $+q$ and $-2q$ at $z = +a$ and $z = +b$, respectively (on the positive $z$-axis). They placed two image charges $-q$ and $+2q$ at $z = -a$ and $z = -b$, respectively (on the negative $z$-axis). They added the electrostatic potential from each charge and wrote the expression for the potential without mentioning where they were finding it. They mentioned that the potential due to a point charge is written as $V = \frac{kq}{r}$, where '$r$' represents the smallest distance between the point charge and surface of the conductor. They also mentioned that since the surface of the conductor is at $z = 0$, '$r$' for the given point charge '$q$' can be written as $r = a - 0 = a$. Eventually, they added the contributions of the potential from charges in the positive z space (region above the conductor) as $V(z > 0) = \frac{1}{4\pi\epsilon_0}\left(\frac{q}{a} - \frac{2q}{b}\right)$. In the notation used by the student, $V(z > 0)$ refers to the potential due to charges present on the positive $z$-axis (it does not refer to the potential in the region $z > 0$, because they never specified the point where they were finding the potential). They wrote the potential due to charges present on the negative $z$-axis as $V(z < 0) = \frac{1}{4\pi\epsilon_0}\left(\frac{q}{a} - \frac{2q}{b}\right)$ in which they incorrectly wrote distances with negative values $\left(V(z < 0) = \frac{1}{4\pi\epsilon_0}\left(\frac{-q}{-a} + \frac{+2q}{-b}\right)\right)$. Then they added the potential due to charges on the positive and negative z axis to obtain the total potential. They wrote $V_{TOT} = 2V(z > 0)$, or the total potential due to all point charges as $V_{TOT} = \frac{2q}{4\pi\epsilon_0}\left[\frac{1}{a} - \frac{2}{b}\right]$. This process of obtaining the potential shows again that the student did not evaluate the expression for potential at any arbitrary point $(x, y, z)$ in space in the region of interest or there was confusion regarding the interpretation of '$r$' in the expression of potential.

Regarding RQ2, to address these difficulties, in the inquiry-based tutorial, students were provided with scaffolding support, e.g., a conversation between hypothetical students illustrating common student difficulties in writing distances correctly. Students were asked to express their agreement or disagreement with each student along with the reasoning. In the following dialogue from the tutorial, Pria clarified the correct approach:

*A point charge 'Q' is placed at a position $z = a$ $(a > 0)$ on the positive z-axis as shown in Figure 7. Consider the following discussion between Emily and Pria.*

**Emily:** *The potential due to the given point charge 'Q' at a point P shown with $(x, y, z)$ coordinates can be written as $V = \frac{kQ}{a}$.*

**Pria:** *This is an incorrect way of writing the distance in the expression for the potential. The potential due to a point charge 'Q' is written as $V = \frac{kQ}{r}$ where '$r$' represents the distance from the point charge to the point P in the space with coordinates $(x, y, z)$ at which we're trying to find the potential.*

Students are provided with scaffolding, e.g., in the form of a partially completed diagram, and are given a hint to help them label the distances correctly as follows:

*How can distance $|\vec{r}|$ be written in terms of coordinates $(x, y, z)$ and $(0,0,a)$? You can write it in Figure 7 shown here.*

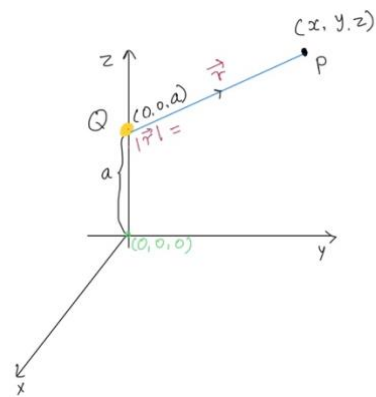

**Figure 7.** A point charge 'Q' placed at a position $z = a$ $(a > 0)$ on the z-axis.

Students were then provided with a checkpoint to strengthen their understanding. We observed that with the scaffolding support offered in the tutorial, interviewed students showed noticeable improvement and learned to write expressions for the distances which were closer to correct. For example, in Q6 of the scaffolded posttest (provided in the Supplementary Material), the first student determined and located the five image charges correctly with correct signs and magnitudes in their diagram. They made sign errors within the expressions for distances for each term and other minor errors in determining the coordinates of the charges mathematically, such that all charges appear to be placed in the third quadrant in the expression shown here (they located those charges correctly in their diagram). They wrote the potential as $V = \frac{1}{4\pi\epsilon_0}\left( \frac{q}{\sqrt{\left(x+\sqrt{3a}\right)^2+(y+a)^2+z^2}} + \frac{-q}{\sqrt{\left(x+\sqrt{3a}\right)^2+(y+2a)^2+z^2}} + \frac{q}{\sqrt{\left(x+2\sqrt{3a}\right)^2+(y+a)^2+z^2}} + \frac{-q}{\sqrt{\left(x+2\sqrt{3a}\right)^2+(y+2a)^2+z^2}} + \frac{q}{\sqrt{\left(x+\sqrt{3a}\right)^2+(y+3a)^2+z^2}} + \frac{-q}{\sqrt{\left(x+\sqrt{3a}\right)^2+(y+2a)^2+z^2}} \right)$. This expression shows that they were now able to understand that the potential should be a function of $(x, y, z)$ and should be determined at an arbitrary point in the region of interest.

Similarly, for one of the tutorial problems, which asked students to find the distances in Cartesian coordinates from each of the two-point charges ($+q$ present on the positive $z$-axis at $z = +d$ and $-q$ present on the negative $z$-axis at $z = -d$) to an arbitrary point $(x, y, z)$ in the space and labeling those distances as $r_1$ and $r_2$, the second student wrote the distances correctly. They wrote $r_1 = \sqrt{x^2 + y^2 + (z-d)^2}$ and $r_2 = \sqrt{x^2 + y^2 + (z+d)^2}$, which shows that these additional scaffolding supports and checkpoints in the tutorial appear to support students develop a clearer understanding for correctly representing distances in their solutions.

### 4.2. Features of the Inquiry-Based Tutorial Providing Scaffolding

Regarding RQ2, we further discuss the features of the tutorial, which helped in improving student learning.

#### 4.2.1. Visual Representations

Recognizing and encouraging the use of diagrammatic representations (Derman et al., 2019) are especially important while solving complex problems in electrostatics. In several cases, even when diagrams were provided or when students were explicitly prompted to draw them, many students drew additional diagrams to support their reasoning. The inquiry-based tutorial's design, offering ample space and opportunities for reflection, seemed to promote this behavior.

#### 4.2.2. Conversations Between Students

The structure of the inquiry-based tutorial includes scripted conversations between fictional students, designed to help students reflect on the types of difficulties commonly observed in interviews. These conversations were grounded in actual difficulties and reasoning errors that students demonstrated while solving problems during interviews. By engaging with these dialogues, students were supported to reflect on their understanding, compare different lines of reasoning, and reconsider their initial assumptions. In several instances, these conversations also included subtle cues or guiding questions that helped students recognize inconsistencies in their thinking and move towards a more accurate solution path. This structure aimed to promote deeper conceptual understanding rather than reliance on memorized procedures.

#### 4.2.3. Embedded Checkpoints

Checkpoints were embedded throughout the tutorial to provide additional scaffolding to support student learning. These checkpoints were strategically placed to prompt students to reflect on the concepts they had just used in the corresponding problems. By encouraging students to pause and consolidate their reasoning, the checkpoints served as moments for self-assessment and conceptual clarification. This structure allows students to recognize errors in their reasoning early and provides opportunities to revise their approach before moving on. Overall, the checkpoints contributed to reinforcing key ideas and supporting deeper engagement with the material.

### 4.2.4. Structure and Flexibility of the Inquiry-Based Tutorial

The inquiry-based tutorial is designed as a sequence of interconnected questions using common difficulties as a guide that gradually builds students' understanding. This progressive structure supports students in developing a more coherent understanding of complex ideas by layering concepts with appropriate scaffolding determined via feedback from students, e.g., in individual interviews. The inquiry-based tutorial follows a structured sequence but is also flexible in its implementation. Instructors can choose to use the entire sequence as in this investigation or select specific sections that are most challenging for students without guidance and relevant to their instructional goals. This adaptability makes the inquiry-based tutorial a useful resource for varied classroom contexts.

### 4.3. Results from the Classroom Implementation

In this section, the results for student performance from the classroom implementation of the scaffolded pre- and posttests across six questions for three instructors are discussed (recall that unscaffolded tests were not given in in-class administration to students). Furthermore, three of the pre-/posttest questions were identical (Q1, Q2 and Q5) and related to the broader issues with method of images while the other three questions were different (Q3, Q4 and Q6) with Test B being more challenging than Test A. These questions that were different on Test A and Test B were related in that they divided a problem that can be solved using MoI into three questions focusing on different aspects of MoI (e.g., the number of image charges needed, boundary conditions and expression of the electric potential). Instructors I1 and I3 administered Test A as pretest and Test B as posttest, whereas I2 administered Test B as pretest and Test A as posttest. Table 1 shows the number of students in each group.

### 4.3.1. Comparison of the Performance of Matched Students and All Students

Since the number of students in I2's class was small, we first compared the results for all students who took the tests (All) with matched students (M) and found them to be consistent. Figure 8 shows the average scores of students in I2's class across six questions for all students and matched students and the standard error and effect sizes (Cohen's $d$, which is typically considered small if $d\sim 0.2$, medium if $d\sim 0.5$ and large if $d\sim 0.8$ (Cohen, 1988)) from pretest to posttest. These students were administered Test B as the pretest and Test A as the posttest. I2Pre_All and I2Post_All refer to all students, and I2Pre_M and I2Post_M refer to matched students for instructor 2. The solid fill is used for Test A and striped pattern is used for Test B. The left pair of bars (striped pattern representing pretest and solid pattern representing posttest) shows the average percentage scores for each question of all students together. The right pair of bars (striped pattern representing pretest and solid pattern representing posttest) show the student average percentage scores of matched students. Numbers on top of bars in boxes represent effect sizes for each pretest and posttest pair. Figure 8 shows the remarkable similarity between the student average percentage scores for all and matched students. No significant differences were found between the average percentage scores of all students and matched students for the class of I2, a pattern that was also consistent with I1 and I3. The observed consistency allowed us to proceed with the analysis using only matched students from all three instructors, hereafter referred to simply as students.

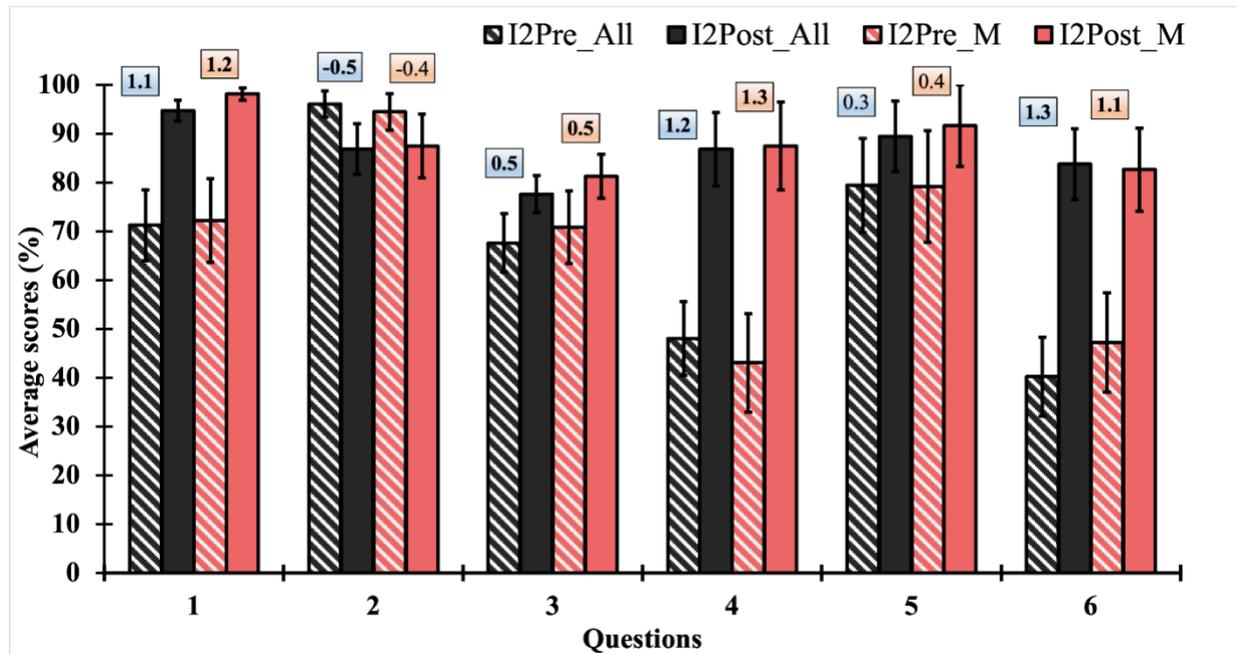

**Figure 8.** Student average percent scores for the method of images pretest and posttest for instructor 2 or I2 for all (I2Pre_All and I2Post_All, left pair) and matched (I2Pre_M and I2Post_M, right pair) students. The standard errors and effect sizes (Cohen's *d*) from pretest to posttest are also shown.

### 4.3.2. Performance of Students for Instructor I1

For I1, Test A was administered as the pretest and the more challenging Test B as the posttest. We note that providing exactly the same test as pretest and posttest could result in students remembering the test questions and only focusing on the answers to those questions instead of learning holistically from the learning tool. Therefore, consistent with (McDermott & Shaffer, 2018), the more challenging Test B was administered as the posttest and Test A as the pretest. We expected that engaging with the tutorial could help students do well even on the challenging problem (with three associated questions) in the posttest (Test B) even when students struggled with the less challenging problem in the pretest (Test A) after a traditional lecture. The student average percentage scores on both the pretest and posttest for I1 are shown in Figure 9. The left pair of bars (blue) with solid fill represents Test A (pretest) and the striped pattern represents Test B (posttest) for I1. Test type is marked for each instructor in Figure 9. Numbers in the boxes refer to the effect sizes (Cohen's *d*) for each pair of pretest and posttest.

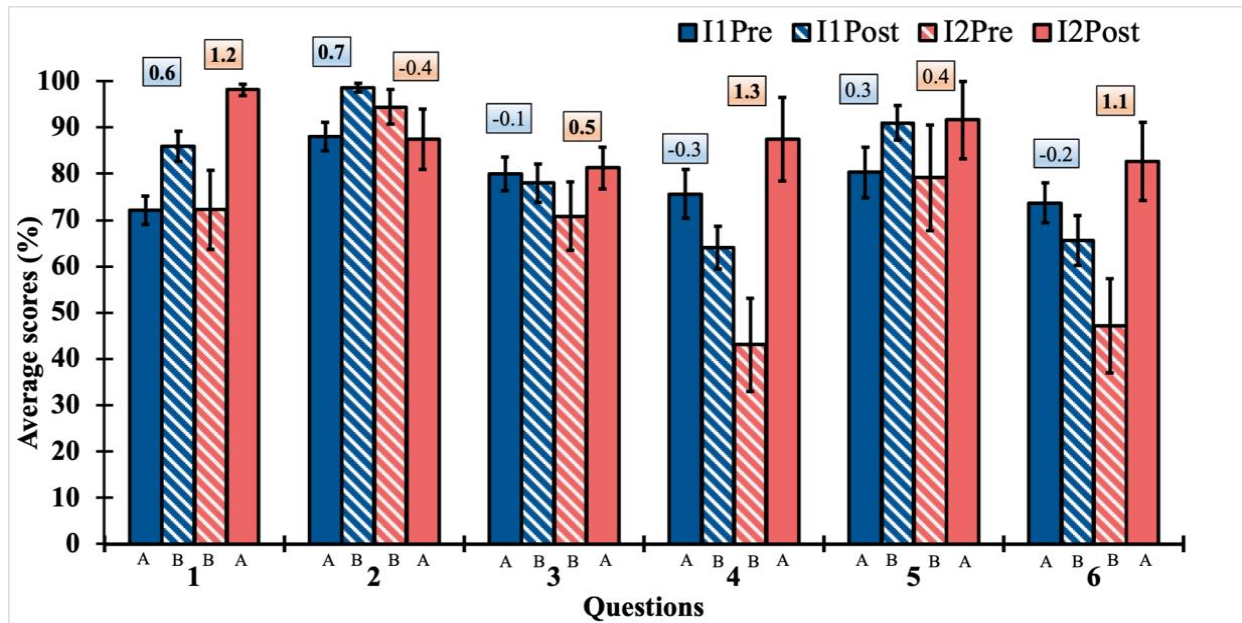

**Figure 9.** Student average percentage scores on the scaffolded pretest and posttest for the method of images for instructor 1 or I1 (I1Pre and I1Post, left pair) and instructor 2 or I2 (I2Pre and I2Post, right pair) for Test A and Test B. The standard errors and effect sizes (Cohen's *d*) from pretest to posttest for each class are also shown.

Q1, Q2, and Q5 were the same in the pretest and posttest in both the test types (Test A and Test B); therefore, these can be directly compared as shown in Figure 9. Q1 asked students to explain the method of images and its relation to the uniqueness theorem. We observe that while students had some understanding of the concept after the traditional lecture, their understanding showed improvement after working through the inquiry-based tutorial. In Q2 of the pretest and posttest, students were asked to circle all of the following conditions which together are necessary and sufficient to properly specify an electrostatic problem with a unique solution. The options were given as: the electric potential is known on the boundaries, the charge density is known outside of the region of interest, and the charge density is known in the region of interest. Since most students performed well on this question prior to the tutorial (ceiling effect), there was limited room for improvement, and the gains were small. Thus, the performance of students on Q2 shows that the test is useful in evaluating prior knowledge of students and that traditional lectures were useful in helping students understand this aspect of the MoI. Q5 asked students to explain if the conductor could be removed in the presence of image charges. Students performed well on this question after engaging with the tutorial. Thus, the tutorial helped in reinforcing important concepts related to the necessary and sufficient conditions for the unique solution and the idea that conductor(s) must be removed when the image charges are introduced. On the three questions (Q3, Q4, and Q6) on which posttest (Test B) was more challenging, performance was lower on the posttest (Test B) than pretest (Test A) due to the higher complexity of posttest questions involving six-fold symmetry.

### 4.3.3. Comparison of Performance of Students for I1 and I2

Figure 9 shows that students of I1 did not perform well on the three challenging questions on the posttest due to complex symmetry of the problem involving six-fold symmetry. Those three questions based on a complex symmetry problem were probing students' understandings in the context of the number of image charges, boundary conditions and expression for the potential. In this design-based comparative research, the researchers decided to switch the pretest and posttest in the second instructor I2's class. In particular, Test B with a more challenging problem associated with three of the questions was given as pretest and Test A was given as posttest in I2's class to investigate how students perform on the challenging problem on Test B administered as pretest (after the traditional lecture). Figure 9 shows student average percentage scores for I2 for each question with the right pair of bars (red) with striped pattern representing Test B (pretest) and solid fill representing Test A (posttest) for I2.

Since the complexity of the problem was higher on Test B, researchers focused on making a direct comparison of performance of students for I1 and I2 on Q3, Q4, and Q6 of Test B. The sample for this analysis is identical to that used in the other analyses presented in the paper. In particular, we compared performance by calculating the effect sizes and normalized gains on Q3, Q4, and Q6 on Test B given as pretest in I2's class with the same Test B given as posttest in I1's class as shown in Table 3. We made these comparisons under the assumption that students and instructors over those consecutive years were comparable. This analysis helped us in evaluating how students, on average, performed on these three challenging questions involving six-fold symmetry on Test B before and after engaging with the tutorial. In addition to Table 3, Figure 9 also shows how students performed from pretest to posttest for I1 and I2 on different test types. Q3 required students to identify the correct number of image charges for a given configuration. Q4 focused specifically on whether students can convert verbal descriptions into mathematical representations, which is often challenging. We found that many students had difficulty with this process, especially when formulating boundary conditions for a given problem. Q6 built upon the concepts assessed in Q3 and Q4. Specifically, students needed to correctly determine the number of image charges (Q3) and describe the mathematical representation of boundary conditions (Q4) to successfully write the expression for the potential in the region of interest (Q6). Therefore, student performance in Q6 was closely linked to their understanding demonstrated in Q3 and Q4.

Table 3 shows that students, on average, typically performed better on Q3, Q4, and Q6 on Test B when they attempted it after the tutorial than after traditional lecture-based instruction, with small to medium (bolded) effect sizes. Table 3 also shows that the normalized gain (Hake, 1998), which is $\frac{\text{post \%} - \text{pre \%}}{100 - \text{pre \%}}$, is also reasonable.

**Table 3.** Comparison of student performance on Test B on three challenging questions (Q3, Q4, and Q6) given in I2's class as pretest and I1's class as posttest (assuming instructors and students in two classes in consecutive years at the same university are comparable). Effect sizes (medium effect sizes are bolded) and normalized gains on Test B (from pretest to posttest) are shown.

| Question | Q3 | Q4 | Q6 |
|---|---|---|---|
| Effect size (Test B) | 0.3 | **0.6** | **0.5** |
| Normalized gain (Test B) | 0.2 | 0.4 | 0.3 |

### 4.3.4. Comparison of Student Performance for I1 and I3 and the Effect of Additional Scaffolding

To further investigate if greater level of improvement in student performance can be achieved on the more challenging questions on Test B, researchers decided to introduce additional scaffolding in the tutorial involving twelve-fold symmetry. This was designed to help students recognize the complex symmetries based on issues researchers observed in in-class performance of students of I1 and I2, as well as interviews with students and discussions with experts. This latest version of the tutorial, which was given to students of instructor I3, was effective for students who were individually interviewed with this latest version. For RQ3 related to this latest version of the tutorial, we hypothesized that additional scaffolding in the tutorial with a twelve-fold symmetry problem would help students perform better on Test B on a six-fold symmetry problem in posttest compared to other students who did not receive this scaffolding in the tutorial. In particular, we expected a more robust improvement in their posttest performance (similar to the improvement in posttest of some of the interviewed students who were interviewed with this latest version).

Figure 10 shows the performance of students on pretest and posttest for I1 and I3. The left pair of bars (blue) with solid fill (pretest—Test A) and the striped pattern (posttest—Test B) represent I1 and the right pair of bars (green) with solid fill (pretest—Test A) and the striped pattern (posttest—Test B) represent I3. Although we expected that after the additional scaffolding in the tutorial, students for I3 will do better on the challenging questions on posttest involving six-fold symmetry compared to I1's students, student performance on the posttest in I3's class was not better. In particular, in the case of I3's class, despite having additional scaffolding in the tutorial, students showed the least improvement. They even scored lowest on the pretest, especially on questions related to identifying the image charges and finding the potential. This outcome suggests that it is possible that there was a lower level of student engagement with the tutorial in I3's class. Some students appeared not to utilize the scaffolding provided in the tutorial, and their lack of engagement may be reflected in their poor posttest performance. In I3's class, the instructor explicitly framed that the inquiry-based tutorial was not a part of current course content, but completing the pretest, tutorial, and posttest, which is a part of a physics education research study, would earn them a small amount of extra credit. This framing by the instructor may have impacted student engagement with the tutorial and their performance.

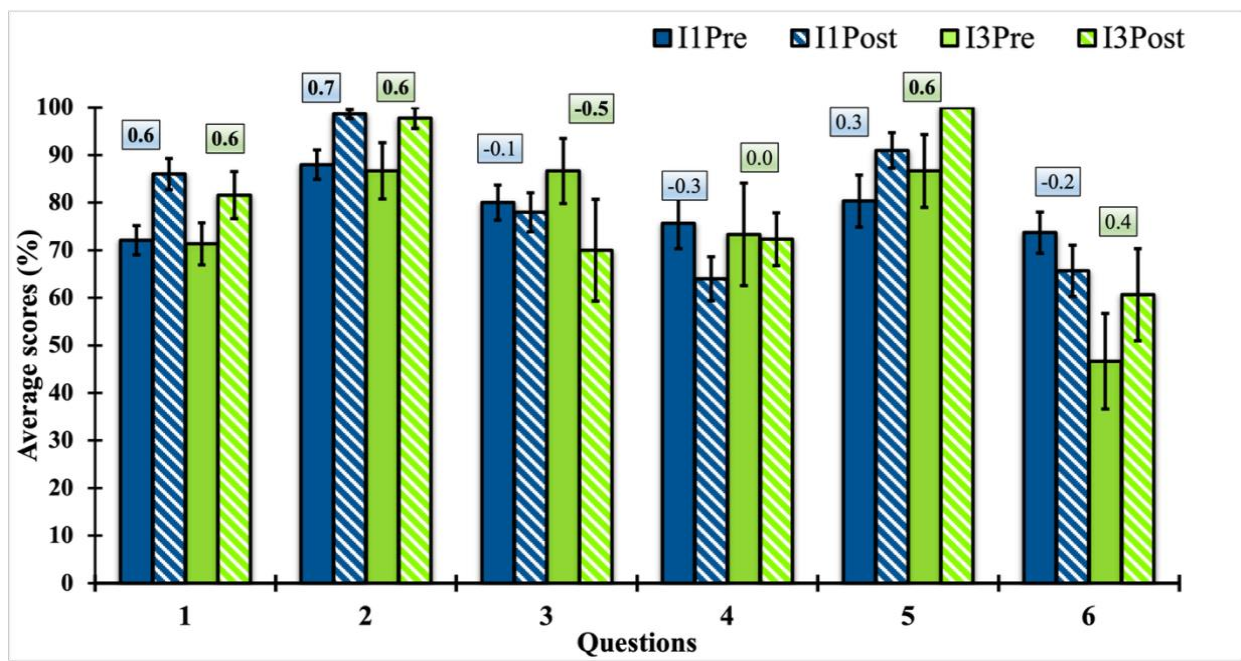

**Figure 10.** Student average percentage scores on the scaffolded pretest and posttest for the method of images for instructor 1 or I1 (I1Pre and I1Post, left pair) and instructor 3 or I3 (I3Pre and I3Post, right pair). The standard errors and effect sizes (Cohen's *d*) from pretest to posttest for each class are also shown.

These findings suggest that while the tutorial could be effective in improving students' understanding of the relevant concepts, factors such as instructor framing of learning activities and the incentives provided to students may have played a significant role in shaping their engagement and, ultimately, their learning outcomes.

### 4.3.5. Key Findings

Figure 11 shows the main findings from this design-based comparative study using a quasi-experimental design and mixed-method triangulation during the iterative process of in-class implementation and refinement of tutorial and corresponding tests over a four-year period. As noted, Test A and Test B were administered as pretest or posttest in which three questions were the same and the other three questions were different.

| Main findings from design-based comparative study |
|---|
| **Instructor I1's class** |
| • On three common questions, posttest performance was better on two questions (Q1, Q5) after tutorial engagement and ceiling effect was observed on one question (Q2) with high pretest score showing traditional lecture was sufficient at least for some concepts.<br>• On the three questions (Q3, Q4, and Q6) on which posttest (Test B) was more challenging, performance was lower on the posttest than pretest (Test A) due to higher complexity of posttest questions. |
| **Instructor I2's class** |
| • Pretest and posttest were switched so that pretest (Test B) was more challenging compared to posttest (Test A). |
| **Comparison of student performance on Test B on three challenging questions (Q3, Q4, and Q6) given in I2's class as pretest and I1's class as posttest (assuming instructors and students in two classes are comparable in consecutive years at same university)** |
| • On challenging Test B questions (Q3, Q4, and Q6), posttest showed moderate improvement compared to pretest. |
| **Instructor I3's class** |
| • Additional scaffolding was added in the tutorial (involving twelve-fold symmetry), and Test A and Test B were given in I3's class in the same order as in I1's class.<br>• Instructor I3 framed the tutorial and tests as not relevant to the course material and part of a research study for which students could earn small amount of extra credit for completion.<br>• Students showed lower engagement and performance on tests and tutorial likely due to instructor framing of the activity and lower incentives. |

**Figure 11.** Main findings from design-based comparative study using quasi-experimental design and mixed-method triangulation during the iterative process of in-class implementation and refinement of the tutorial and corresponding tests over a four-year period.

## 5. Broader Discussions

This study provides valuable insight into ongoing work in physics education research aimed at improving learning of students in upper-level electricity and magnetism courses. Using an iterative multi-year design-based research study involving the development and validation of an inquiry-based tutorial on the MoI, we identified common student difficulties and assessed how the inquiry-based sequences in the tutorial support students in solving MoI problems. In addition, we examined how the instructors' framing of the learning tools to students and the incentives offered by them influenced the engagement and performance of students. Apart from conducting individual interviews throughout, we employed multi-instructor implementation over a four-year period and mixed-method triangulation.

Physics is a hierarchical discipline, and it requires students to effectively integrate mathematics and physics well. If appropriate support is not provided, student difficulties can persist and continue to affect learning even at the advanced levels. Some of the common difficulties found via this research, e.g., related to number and location of image charges, are specifically in the context of MoI. However, other difficulties, e.g., related to grounded conductors or difficulties with measuring distances for finding the potential are similar to those found in other studies. For example, a prior study in the context of conductors and insulators (Santana et al., 2023) shows that students at the introductory level had many difficulties related to grounded conductors including the role of grounding similar to those found in this study. In particular, introductory students sometimes claimed that grounded spherical conductor will maintain equal amount of positive and negative charges (Santana et al., 2023) similar to the advanced students in this study. Prior research by Harrington (Harrington, 1995) suggests that students often interpret 'grounding' as 'neutral', reflecting a common difficulty in understanding this concept, similar to the responses observed in our study. Research (Bilak & Singh, 2007) also shows that students have difficulty with charges on the conductor in the presence of electrical induction and grounding, similar to the difficulties observed in this study when students drew charge distribution on the grounded conducting plane(s) placed near a point charge. Similarly, a prior study in the context of electric field shows that students struggle with the correct distances to use in the expression for electric field (Li et al., 2023) similar to the struggle upper-level students had with correct expressions for distances while finding the potential in this study.

The comparison of student performance on pretest and posttest suggests that the tutorial helped in student understanding of the method of images at least on some questions. While the difficulty level of the pretest and posttest was not the same, after engaging with the inquiry-based tutorial, students demonstrated improvements in their ability to find and write expressions for electrostatic potential in the region of interest. Also, since the difficulty levels of pre- and posttests were different, we also calculated effect sizes and normalized gains for students in different instructors' classes to compare average improvements on three of the problems on the challenging test (Test B) assuming students and instructors across different years are comparable. These findings show that the MoI tutorial is helpful for some of the concepts students were tested on. Interviews also showed that students benefited from the inquiry-based sequences in the tutorial.

The inquiry-based tutorial also encouraged the use of effective problem-solving strategies, such as drawing diagrams. We find that many students independently drew diagrams to assist their reasoning, despite it not being required. This behavior was likely reinforced by the tutorial's design, which included space for sketching and opportunities to reflect. These findings suggest that explicitly encouraging and rewarding effective practices may improve students' ability to solve problems.

Furthermore, an important finding of this study is that the instructor's framing and incentives to engage effectively with inquiry-based instructional tasks such as the tutorial and accompanying tests can impact student engagement and performance. In particular, the way an instructor frames the value of an inquiry-based instructional task and provides incentives to engage meaningfully with it can impact student engagement and learning. Previous research (B. R. Brown et al., 2016) suggests that the engagement and performance of students are closely tied to how learning activities are introduced and incentivized. When students view learning tasks as meaningful and connected to the course goals, they are more likely to engage with them thoughtfully. On the other hand, if an inquiry-based instructional activity is framed as unrelated to the course goals and offered without incentives for engagement and learning (Kashyap et al., 2025), all students may not invest effort to engage meaningfully. We observed that in one of the classes, the inquiry-based tutorial and tests were described as part of a research activity and the instructor explicitly told students that the material was not directly relevant to the current course for future testing and exam perspective. A small amount of extra credit was provided to students for participation in the inquiry-based learning sequences in the tutorial, regardless of correctness of their work. In this context, some students appear not to have engaged deeply with the material and did not perform well on the tests. These differences suggest that how instructors frame the purpose and importance of inquiry-based learning activities can influence students' motivation to engage meaningfully and their performance.

## 6. Theoretical, Practical, and Policy Implications

Overall, this study provides important insights for instructors and researchers investigating student difficulties, and developing, validating and implementing research-based learning tools aiming to support student learning. One major finding is that similar to the common difficulties found among students in introductory physics courses, we find that advanced students also have similar common difficulties with physics concepts. In particular, our findings indicate that some of the difficulties discussed in this research are consistent with those reported in previous studies in physics courses. Notably, even as students advance in their studies, certain conceptual challenges in physics remain.

This iterative multi-year design-based research study with mixed-method triangulation underscores the challenges involved in such studies that educators and researchers can benefit from. We made iterative changes in the tutorial based on student interviews, feedback from experts and classroom data during the entire process of development and validation of the tutorial and corresponding tests. Since the in-class engagement and performance of students is driven by factors where researchers do not have complete control compared to interviews where students are asked to think-aloud as they solve problems and observed in real time, the outcomes for in-class and out-of-class can be different. The process of observing student problem-solving during interviews also highlighted instructional strategies that can enhance problem-solving approaches. For example, encouraging students to draw multiple diagrams and use different representations appears to have supported an improved approach in problem-solving. Although the overall impact of the in-class implementation of the tutorial may be influenced by many factors, the scaffolding embedded in the tutorial is grounded in observed student common difficulties.

The study highlights the challenges involved in making the pre/posttest different. Researchers are often concerned about making all of the pre/posttest questions the same in case students remember the questions from the pretest and only learn the answers to those questions without developing a holistic understanding of all of the concepts involved. For example, in physics education, some researchers (McDermott & Shaffer, 2018) have given a more challenging posttest after students have engaged with a physics tutorial. In the research described here, half of the pre/posttest questions were the same and other half of the questions were more challenging in Test B that was given as a posttest by instructor I1 over the first two years. Since this more challenging test masked student learning from the tutorial in instructor I1's class, the researchers switched the order of tests in instructor I2's class. Then the comparison of Test B given as a pretest in instructor I2's class with the same test given as a posttest in instructor I1's class showed moderate improvement on that challenging test (Test B) after engaging with the tutorial. However, this analysis is based on assuming that both instructors were comparable in traditional lectures and students in two classes were comparable.

We find that the latest version of the tutorial and the corresponding tests administered in instructor I3's class did not show the same positive outcome we found in the few individual student interviews we conducted with that latest version. Since the third instructor had framed for students that the tutorial and the corresponding tests were not relevant for the course and they would achieve a small amount of extra credit, it is likely that some of the students did not engage with the tutorial or tests in a meaningful way, which was reflected in their overall performance. A policy implication is that in the professional development of instructors and teaching assistants, who are early-career educators, they should be asked to reflect on the importance of instructional framing on student motivation, engagement, and learning. In particular, it is important for instructors to clearly convey the positive value of instructional tasks and incentivize them, e.g., via grade incentives, so that students would view the activities as meaningful and rewarding. These mindful strategies may enhance student engagement and learning.

## 7. Limitations and Future Directions

One limitation of this study is that it is a quasi-experimental comparative study involving an iterative design process and uses mixed-method triangulations. Also, the difficulty levels of pre- and posttests were different; therefore, we also calculated effect sizes and normalized gains for students in different instructors' classes to compare improvements on the same challenging test (Test B) assuming students and instructors across different years are comparable. These findings show that after engaging with the tutorial, student performance was moderately improved even on the more challenging Test B; however, it is not possible to make strict comparisons across instructors in different years. Also, the study involved small sample sizes for two of the instructors which further limits the statistical power of quantitative analyses and hence the generalizability of our findings. Some of the other limitations of this investigation are that all data are from a single institution and the number of students in each class was small, which limits generalizability. Also, each of the three instructors' teaching approaches may have influenced students' pretest performances before engaging with the inquiry-based tutorial.

Future investigations should explore the generalizability of results discussed here by conducting similar investigations with the tutorial at other institutions of different types, e.g., small colleges and large universities, across different countries. Moreover, due to time limitations and instructors' choices, the tutorial was given as homework across all three classes, potentially restricting collaborative engagement. The tutorial could be adapted for in-class use in the future, to encourage discussion and collaborative learning and reasoning, providing opportunities for students to work together in small groups (Crouch et al., 2007; Lorenzo et al., 2006; Miller et al., 2015; Olschewski et al., 2023; Pozzi et al., 2023). Such implementations are likely to lead to deeper engagement and more robust learning outcomes from inquiry-based sequences in the tutorial. Moreover, after the concepts are introduced through lecture-based instruction, inquiry-based tutorial questions can be restructured for use in think-pair-share discussions or as clicker questions (Laal, 2013; Laal & Ghodsi, 2012; Le et al., 2018; Mazur, 2009; Miller et al., 2014; T. Nokes-Malach et al., 2015; Watkins & Mazur, 2013) to help students learn effectively with peers.

**Supplementary Materials:** The following supporting information can be downloaded at *https://www.mdpi.com/article/10.3390/educsci16040594/s1* , Test A, Tutorial, Test B.

**Author Contributions:** Conceptualization, C.S.; methodology, C.S. and R.P.D.; validation, J.S.K., C.S. and R.P.D.; formal Analysis, J.S.K.; investigation, J.S.K., C.S. and R.P.D.; data curation J.S.K., R.P.D.; writing—original draft preparation, J.S.K. and C.S.; writing—review and editing, J.S.K., R.P.D. and C.S.; visualization, J.S.K.; supervision, C.S.; project administration, C.S. All authors have read and agreed to the published version of the manuscript.

**Funding:** This research received no external funding.

**Institutional Review Board Statement:** The study was conducted in accordance with the Declaration of Helsinki, and approved by University of Pittsburgh Institutional Review Board (protocol code PRO15070212 and date of approval: 20 July 2015).

**Informed Consent Statement:** All interview participants provided individual consent.

**Data Availability Statement:** The raw data presented in this study are not available as per institutional IRB policy.

**Acknowledgments:** We thank Patrick Grugan for helping us in developing the initial version of the tutorial. We also thank the students and professors in the department who helped by giving constructive feedback and suggestions to make the tutorial better.

**Conflicts of Interest:** The authors declare no conflicts of interest.

## Appendix A

This appendix includes the latest version of Test A without spaces after each question that students were provided to answer the questions. Test B had a similar format but did not have the additional problem Q7 (included in Supplementary Material). The Supplementary Material includes the latest version of the entire tutorial, pretest, and posttest.

1.  Explain in your own words the idea behind the method of images, what it accomplishes, and its relationship to the uniqueness theorem.

2.  Circle all of the following conditions that together are necessary and sufficient to properly specify an electrostatics problem with a unique solution.

    I.   The electric potential is known on the boundaries.
    II.  The charge density is known outside of the region of interest.
    III. The charge density is known in the region of interest.

3.  Consider a grounded conducting sheet (spanning across the $xy$ plane) whose top surface is the $z = 0$ plane. The region $z > 0$ is vacuum. Two-point charges are placed on the $z$-axis: charge $+q$ at $z = a$ and charge $-2q$ at $z = b$ $(a > 0, b > 0$ and $b > a)$. How many image charges are required to find the potential in the region of interest $(z > 0)$ using the method of images? Explain.

    (a) 1 image charge
    (b) 2 image charges
    (c) 3 image charges
    (d) 4 image charges

4.  Write down all of the relevant boundary conditions for the problem posed in Q3 mathematically.

5.  Many method of images problems deal with charges near grounded conductors, e.g., Q3. Consider the following statements from two students about Q3:

    •   Emily: Once we have figured out the image charges for this problem, then we have a problem that only involves point charges, which we will use to find the potential. The conductor is not present in this new problem with image charges.
    •   Judy: I disagree. We cannot get rid of the conductor in the new problem with image charges when we calculate the potential using the image charges.

With which student do you agree? Explain.

6.  Using the information in Q3, solve for the potential in the region of interest.

7.  Consider two grounded conducting planes as shown in Figure A1, both perpendicular to the $x$-$y$ plane, intersecting at an angle of 30°. One plane is at y = 0 and the other makes a 30° angle with respect to it. A point charge $+q$ is placed at a point along the line bisecting the region at 15° in the region of interest. How many image charges are required to solve the problem?

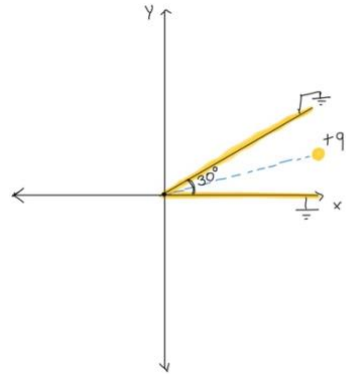

**Figure A1.** A point charge 'q' present in the region of interest between two grounded conducting planes placed at 30°..

*Appendix A.1. Rubric*

Rubric for the latest versions of Test A and Test B are shown in Table A1 and Table A2. For Q6, the rubric is somewhat different for Test A and Test B since the problems were different.

**Table A1**. Rubric for the latest version of Test A.

| Question | Deduction % | Explanation |
|---|---|---|
| Q1 | −0% | Correct. |
| | −22% | Mostly correct. |
| | −11% | Mostly correct, with a minor error and/or lack of clarity on some points. |
| | −33% | Partially correct. |
| | −100% | No serious attempt to answer the question. |
| | −67% | Stated uniqueness theorem, but nothing more or wrote some relevant things but the explanation is far from complete. |
| Q2 | −0% | Correct. |
| | −50% | Choice III should be circled. |
| | −50% | Choice I should be circled. |
| | −100% | Choice II is not correct. |
| Q3 | −0% | Correct. |
| | −50% | Incorrect number of image charges (should be 2). |
| | −50% | No explanation. |
| | −25% | Incorrect explanation. |
| Q4 | −0% | Correct. |
| | −17% | Extra boundary condition. |
| | −33% | One incorrect or missing boundary condition. |
| | −50% | Two incorrect boundary conditions. |
| | −67% | Miscellaneous deduction. |
| | −17% | Minor deduction. |
| | −100% | No answer or incorrect answer. |
| | −50% | Incorrect specification of the boundary: $V(x, y, 0) = 0$ is correct. |
| Q5 | −0% | Correct. |
| | −50% | Incorrect or indecipherable explanation. |
| | −50% | Emily is correct. |
| | −33% | Miscellaneous deduction. |
| Q6 | −0% | Correct. |
| | −17% | Sign errors. |

| −25% | Multiple errors in positions of the charges. |
|------|-----------------------------------------------|
| −100% | No attempt. |
| −8% | One pair of terms has a sign error. |
| −17% | Did not specify the positions of the charges analytically. |
| −33% | Miscellaneous errors. |
| −8% | Miscellaneous minor error. |
| −25% | Miscellaneous deduction. |

**Table A2**. Rubric for the latest version of Test B.

| Question | Deduction % | Explanation |
|----------|-------------|-------------|
| Q1 | −0% | Correct. |
| | −22% | Mostly correct. |
| | −11% | Mostly correct, with a minor error and/or lack of clarity on some points. |
| | −33% | Partially correct. |
| | −100% | No serious attempt to answer the question. |
| | −67% | Stated uniqueness theorem, but nothing more or wrote some relevant things but the explanation is far from complete. |
| Q2 | −0% | Correct. |
| | −50% | Choice III should be circled. |
| | −50% | Choice I should be circled. |
| | −100% | Choice II is not correct. |
| Q3 | −0% | Correct. |
| | −50% | Incorrect number of image charges (should be 5). |
| | −50% | No explanation. |
| | −25% | Incorrect explanation. |
| Q4 | −0% | Correct. |
| | −17% | Extra boundary condition. |
| | −33% | One incorrect or missing boundary condition. |
| | −50% | Two incorrect boundary conditions. |
| | −67% | Miscellaneous deduction. |
| | −17% | Minor deduction. |
| | −100% | No answer. |
| Q5 | −0% | Correct. |
| | −50% | Incorrect or indecipherable explanation. |
| | −50% | Emily is correct. |
| | −33% | Miscellaneous deduction. |
| Q6 | −0% | Correct. |
| | −17% | Sign errors. |
| | −25% | Multiple errors in positions of the charges. |
| | −100% | No attempt. |
| | −17% | One pair of terms has a sign error. |
| | −33% | Did not specify the positions of the charges analytically. |
| | −33% | Miscellaneous errors. |
| | −17% | Miscellaneous minor error. |